# H-loop Histidine Catalyzes ATP Hydrolysis in the *E. coli* ABC-Transporter HlyB

Yan Zhou, Pedro Ojeda-May, and Jingzhi Pu*

Department of Chemistry and Chemical Biology, Indiana University-Purdue University Indianapolis, 402 N. Blackford St., Indianapolis, IN 46202.

**ABSTRACT:** Adenosine triphosphate (ATP)-binding cassette (ABC) transporters form a family of molecular motor proteins that couple ATP hydrolysis to substrate translocation across cell membranes. Each nucleotide binding domain of ABC-transporters contains a highly conserved H-loop Histidine residue, whose precise mechanistic role to motor functions has remained elusive. By using combined quantum mechanical and molecular mechanical calculations, we showed that the conserved H-loop residue H662 in *E. coli* HlyB, a bacterial ABC transporter, can act first as a general acid and then as a general base to facilitate proton transfers in ATP hydrolysis. Without the assistance of H662, direct proton transfer from the lytic water to ATP results in a greatly elevated barrier height. Our findings suggest that the essential function of the H-loop residue H662 is to provide a "chemical linchpin" that shuttles protons between reactants through a relay mechanism, thereby catalyzing ATP hydrolysis in HlyB.

Haemolysin B (HlyB) is the inner membrane component of the type I protein secretion apparatus that transports the 104 kD pore-forming toxin Haemolysin A (HlyA) out of gram-negative bacterial cells.[1] It is well established that HlyB functions as an ATP-binding cassette (ABC)-transporter[2] by converting chemical free energy derived from ATP hydrolysis to mechanical work required for substrate translocation. Like many members in the ABC transporter family, HlyB contains two cytosolic nucleotide binding domains (NBDs) and two transmembrane domains (TMDs). As the molecular motor component of the transporter, HlyB-NBDs perform two hallmark functions. On the one hand, the induced conformational changes on HlyB-NBDs, as a result of ATP binding, hydrolysis, and/or product release, energize the translocation process by mechanically coupling to TMDs for substrate passage.[3] On the other hand, HlyB-NBDs work as an enzyme ATPase that catalyzes ATP hydrolysis. In the case of *E. coli* HylB, its NBD hydrolyzes ATP at $k_{cat}$=0.2 s$^{-1}$,[3] corresponding to a rate acceleration of more than 6 orders of magnitude (at room temperature) compared to the solution phase reaction ($k = 8\times10^{-8}$ s$^{-1}$).[4] As the enzymatic activity of the motor is an essential element in achieving efficient mechano-chemical coupling that leads to transport, a thorough understanding of the catalytic mechanism of HlyB-NBDs is a key step in describing how the transporter works.

Despite the large body of data from biochemical studies[3,5-7], sequence analyses,[8] and structural characterizations,[3,6,7] the precise mechanism of ATP hydrolysis in HlyB regarding the roles of active site residues remains unavailable. It is known that the active site of HlyB-NBD contains a collection of highly conserved sequence motifs, including the Walker A motif (or P-loop: $^{502}$GXXXXGKST$^{510}$, where X denotes any amino acid; *E. coli* HlyB sequence numbers will be used throughout this paper) and the Walker B motif ($^{626}\phi\phi\phi\phi$D$^{630}$, where $\phi$ is mostly a hydrophobic residue), the C-loop ($^{606}$LSGGQ$^{610}$, from the opposite monomer), which is signature to the ABC family, the Glu (E631) located immediately after the Walker B motif, and the H-loop that contains an His (H662)[3] (see Supporting Information SI.1 for a schematic representation of key interactions involved in the active site). The functional form of HylB-NBDs is a dimer, which adopts a "head-to-tail" arrangement such that each of the two ATP molecules bound at the dimer interface is sandwiched between the P-loop of one subunit and the C-loop of the other subunit.

The roles of the Walker A and B motifs are relatively well established with the Walker A known responsible for nucleotide binding and the Walker B for Mg$^{2+}$ binding. Unlike the Walker motifs, the contribution of the H-loop His (H662) to catalysis has remained elusive. The conserved H-loop H662 is located in the switch II region of HlyB-NBDs.[9] Sequence and structural comparisons of ABC-transporters and helicases have suggested that the H-loop His may act as a sensor for the γ-phosphate of ATP,[8] therefore reminiscent of the conserved Gln in RecA.[10] Mutagenesis studies showed that the mutation of this H-loop His to Ala (H662A) reduces the ATPase activity of the NBD in HlyB to background levels (<0.1% residual ATPase activity).[3,7] The involvement of the H-loop in hydrogen bonding interactions with the P-loop within the same NBD and with the D-loop ($^{634}$SALD$^{637}$) in the opposite NBD, as revealed in crystal structures of HlyB, suggests roles of H662 in ATP binding and in conformational signaling across the NBD dimer interface.[3] Based on a crystal structure of ATP/Mg$^{2+}$ bound dimeric NBDs of HlyB that contains mutation at the position of H662, Schmitt and co-workers proposed that H662 and E631 form a catalytic dyad, in which the H662 acts as a "linchpin" that holds other active site residues at their catalytically competent conformations.[3] The direct participation of the H-loop H662 in the enzyme mechanism, however, has not been discussed in the literature. Investigation of H662's explicit role in the mechanism of ATP hydrolysis is the focus of the present study.

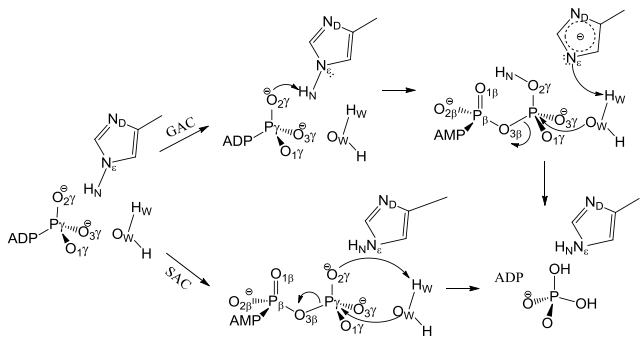

Scheme 1. Proposed mechanisms for ATP hydrolysis in HlyB-NBDs. In the substrate-assisted catalysis (SAC) mechanism, a proton is directly transferred from the lytic water to ATP, whereas in the general acid catalysis (GAC) mechanism, proton transfers are mediated by the H-loop His (H662) via a relay mechanism.

Here we report computer simulations of ATP hydrolysis catalyzed by HlyB-NBDs based on a combined quantum mechanical and molecular mechanical (QM/MM) approach,[11] with which we directly compared two possible reaction mechanisms as depicted in Scheme 1. In the first mechanism (referred to as the GAC mechanism), H662 initially serves as a general acid by donating its proton at the $N_\epsilon$ position to the $\gamma$-phosphate of ATP and subsequently accepts a proton from the lytic water. Since H662 returns to its original protonation state at the end of the catalytic cycle, the net reaction in the GAC mechanism is that a proton is transferred from the lytic water to ATP under the relay of H662. In the second mechanism, we examined the substrate-assisted catalysis mechanism (referred to as the SAC mechanism),[3] where the lytic water directly transfers its proton to one of the $\gamma$-phosphate oxygens. As H662 does not directly participate in the chemical steps in SAC, it would more likely provide structural factors that stabilize the active site configuration, as suggested in the mechanical "linchpin" proposal.[3]

Two crystal structures of *E. coli* HlyB (PDB: 1XEF[3] and 2FGK[6]) were used in present work. The combination of the two structures is necessary to model the active site for the NBD dimer (see SI.2 for details). The NBD dimer was solvated in a water sphere with a radius of 30.4 Å (Figure 1a). A combined QM/MM[12] method in the CHARMM program[13] was employed to obtain the two-dimensional potential energy surfaces (PESs) for both the GAC and the SAC mechanisms. The simulation system was centered at the active site of chain B of the 1XEF structure and partitioned into two regions. The QM region, described by the semiempirical Austin Model 1 (AM1)[14] (see SI.3 for justification), consists of the three phosphate groups of ATP, side chains of S504, K508, and H662 in chain B, the side chain of S607 in chain A, the lytic water ($W_L$), and five boundary carbon atoms that are treated by the generalized hybrid orbital (GHO) method[15] (Figure 1b). The MM region, described by the CAHRMM22 force field[16] with CMAP,[17] contains the rest of the system.

A cutoff distance of 16.0 Å was used for non-bonded interactions with a force switch function in the region of 15-16 Å. A stochastic boundary condition[18] treatment was used, such that a sphere of 24 Å was selected as the reaction region, in which atoms were simulated by Newtonian dynamics, and atoms located in the buffer region, defined as the layer between 24 and 30.4 Å from the center of the solvated sphere, were treated by Langevin dynamics.[19] Bonds involving hydrogens in the MM region were constrained using the SHAKE algorithm[20] during simulations. The system was first heated up gradually to 298 K during 60 ps, followed by an equilibration simulation of 100 ps. After the system was equilibrated, PESs were obtained by scanning each reaction coordinate (RC) at a step size of 0.1 Å with the RESD keyword in CHARMM[13] (see SI.4 for additional simulation details).

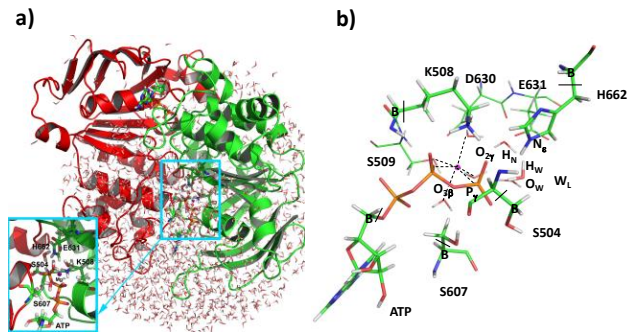

Figure 1. (a) Stochastic boundary setup of the simulation system, where the active site is solvated by a water sphere of 30.4 Å, centered at the $\gamma$-phosphorous atom of ATP. (b) Close-up of the QM region, which consists of three phosphate groups of ATP, side chains of S504, K508, and H662 in chain B of HlyB-NBD, the side chain of S607 in chain A, the lytic water, and five GHO boundary atoms. The components shown in the licorice representation are originally located in the 1XEF structure, except side chains of H662 and E631, and the lytic water, which were modeled from the 2FGK structure. This figure was generated by Pymol.

Two-dimensional PESs for both mechanisms were obtained (Figure 2). For the GAC mechanism (Figure 2a), the abscissa (RC1) represents a proton relay reaction coordinate involving H662 [RC1$^{GAC}$ = ($H_N$–$N_{\epsilon,H662}$ – $H_N$–$O_{2\gamma}$) + ($O_W$–$H_W$ – $H_W$–$N_{\epsilon,H662}$); here the atom pair connected by a '-' denotes the bond distance]. For the SAC mechanism (Figure 2b), RC1 describes a direct proton transfer (RC1$^{SAC}$ = $O_W$–$H_W$ – $H_W$–$O_{2\gamma}$). The ordinate (RC2) in both mechanisms uses the antisymmetric combination of the bond distances of $O_{3\beta}$–$P_\gamma$ and $P_\gamma$–$O_W$ to describe the phosphoryl transfer reaction during ATP hydrolysis, i.e., RC2$^{GAC}$= RC2$^{SAC}$ = $O_{3\beta}$–$P_\gamma$ – $P_\gamma$–$O_W$. Below we describe the major features of the PESs for the two reaction mechanisms. More detailed discussion of geometries of stationary points in Figure 2 is available in SI (see SI.5 & Tables S1&S2).

The PES of the GAC mechanism (Figure 2a) reveals a four-step process for ATP hydrolysis. In the first step, $W_L$ moves close to ATP to form a stable reactant complex (Min2) through a transition state TS1 (see also SI.5 Table S2). In the second step, a proton is transferred from $N_\epsilon$ position of H662 to $O_{2\gamma}$ of ATP through a transition state TS2 that corresponds to a barrier height of 15.0 kcal/mol, leading to an intermediate (Min3) (see SI.5). The proton transfer from H662 to ATP is decoupled from phosphoryl transfer, as the $P_\gamma$–$O_{3\beta}$ bond does not change significantly in Min2, TS2, and Min3 (see SI.5). The third step consists of two concerted processes in which the proton transfer of $H_W$ from the lytic water oxygen $O_W$ to $N_\epsilon$ of H662 is coupled to the $P_\gamma$–$O_W$ bond formation (Scheme 1). Being the rate limiting step in the GAC mechanism, the third step goes through a transition state TS3 that is associative in nature and gives rise to an overall barrier height of 22.1 kcal/mol, resulting in an intermediate Min4. In both TS3 and Min4, $P_\gamma$ has a trigonal bipyramidal bonding pattern. In the last step, cleavage of the $P_\gamma$–$O_{3\beta}$ bond leads to formation of the final product complex (Min5).

The PES obtained for the SAC mechanism (Figure 2b) also displays a stepwise pattern. The SAC reaction route is initiated with formation of a reactant complex (Min2) between $W_L$ and ATP. The second step in SAC is rate limiting and involves two concerted processes in which proton transfer between the lytic water and $O_{2\gamma}$ of ATP is coupled to the $P_\gamma$-$O_W$ bond formation, going through a trigonal bipyramidal transition state TS2. The completion of the rate-limiting step leads to an intermediate (Min3). The third step from Min3 to Min4 corresponds to a structural relaxation of hydrogen bonding interactions involving H662 and E631, followed by the final step in which the $P_\gamma$-$O_{3\beta}$ bond is cleaved (see SI.5 for details).

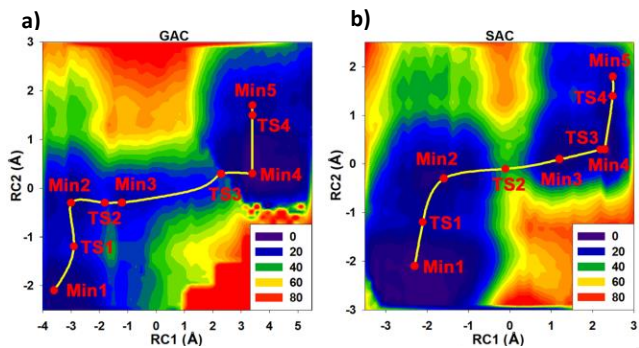

**a)** GAC
**b)** SAC

Figure 2. Two-dimensional potential energy surfaces for the general acid catalysis (GAC, 2a) and the substrate-assisted catalysis (SAC, 2b) in the HlyB-NBDs mediated ATP-hydrolysis. For the GAC mechanism, the abscissa (RC1) represents a proton-relay reaction coordinate which is negative as protons are bonded with lytic water and at $N_\varepsilon$ position of H662; zero as the H662 proton is partially transferred to $\gamma$-phosphate oxygen $O_{2\gamma}$ of ATP and the lytic water proton is partially transferred to H662, and positive as the proton relay is completed. For the SAC mechanism, RC1 represents a direct proton transfer from the lytic water to ATP. The ordinate (RC2) in both mechanisms is the antisymmetric combination of the $O_{3\beta}$-$P_\gamma$ and $P_\gamma$-$O_W$ bond distances, in describing the phosphoryl transfer reaction during ATP hydrolysis. In both figures, distances are in Å, and energies are in kcal/mol. Reaction paths identified here as solid yellow curves (see SI.6 for movies)

The energies of the stationary points in the GAC and the SAC mechanisms were compared in Figure 3. The highest barrier for the SAC mechanism is 32.1 kcal/mol (SAC:TS2); the fact that this barrier height is close to that of ATP hydrolysis in aqueous solution (27.4 kcal/mol, estimated based on a rate constant of $k=8\times10^{-8}$ s$^{-1}$)[4] suggests that SAC is unlikely a viable catalytic mechanism in HlyB. In contrast, in the GAC mechanism, which allows H662 to assist proton transfers, the highest barrier is only 22.1 kcal/mol (GAC:TS3), in good agreement with a free energy barrier of 18.6 kcal/mol estimated from experimental $k_{cat}$ of 0.2 s$^{-1}$ for HlyB.[3] The reduced barrier in the GAC mechanism strongly suggests that H662 is capable of catalyzing the hydrolysis of ATP in HlyB-NBDs by lowering the barrier height of the rate limiting step by about 10 kcal/mol compared to that in the SAC mechanism.

The rate limiting step in the SAC mechanism is the step of transferring $H_W$ to $O_{2\gamma}$ (coupled with $P_\gamma$-$O_W$ bond formation), through a four-center transition state TS2. The involvement of H662 converts this single proton transfer step into two consecutive steps that relay two protons, therefore the geometrically strained four-center reaction is avoided[21] and the activation energy required for the protonation of $O_{2\gamma}$ is reduced by 17 kcal/mol [$\Delta E$(TS2$^{SAC}$- TS2$^{GAC}$)]. The current results suggest that the rate-limiting step in the GAC mechanism, no longer being the protonation of $O_{2\gamma}$, shifts to the step that cou-ples proton transfer of $H_W$ to H662 with the $P_\gamma$-$O_W$ bond formation (TS3) (see SI.6 for movies of reaction steps).

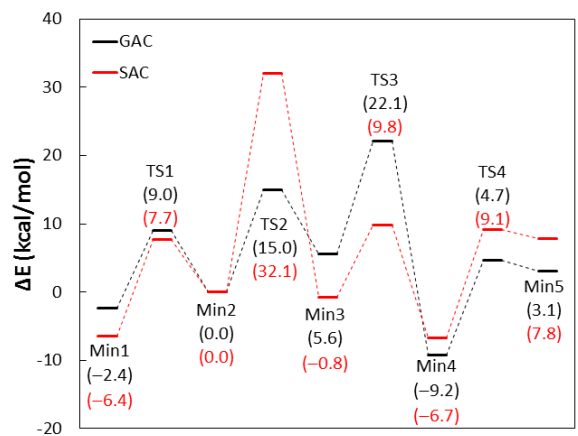

Figure 3. Comparison of energy profiles of the GAC (black) and the SAC (red) mechanisms.

In this Communication, we have shown computationally how the H-loop residue H662 in HlyB-NBDs may catalyze ATP hydrolysis by providing a chemical "linchpin" for assisted proton transfers. In addition to these structure-based simulation results, several lines of biochemical evidence also support the H662 mediated GAC mechanism we proposed here. First, the greatly reduced ATPase activities (<0.1% residual ATPase activity) observed in the H662A mutant of HlyB[3,7] can be rationalized as such substitution disables the GAC proton relay pathway due to lack of exchangeable proton in Ala. Similar reduction in ATPase activities upon single mutation of the H-loop His have also been reported for other ABC-transporters, including the histidine permease HisP (H211D, H211R, and H211Y, <2% residual ATPase activity)[22] and the Maltose transporter MalK (H192R, <2% ATPase activity).[23,24]

The GAC reaction mechanism is also in accord with the pH dependence of the ATPase activity reported for HlyB. The wild-type HlyB displays maximal ATPase activities when pH = 7,[3] at which H662 is likely to be found neutral with its imidazole ring adopting a singly protonated state, activating the GAC mechanism. When pH << 7, H662 becomes doubly protonated H662H$^+$, and ATP is also likely protonated as the pK$_a$ value of ATPH$^{3-}$ (the terminal phosphate) is about 6.5.[25] As a result, the tendency of ATP to remove a proton from H662 diminishes, and H662 in turn loses its catalytic capability of abstracting a proton from the lytic water to activate the nucleophilic attack step in hydrolysis. When pH >> 7, the populated H662$^-$ is expected to deactivate the GAC mechanism, as the anionic form of H-loop His is unable to provide exchangeable proton.

For HlyB-NBDs, the conserved E631 has been proposed as a putative catalytic base which deprotonates the lytic water $W_L$. Interestingly, the pH dependence of ATPase activity in the E631Q mutant of HlyB is almost identical to that of the wild-type enzyme.[26,27] This observation led to a proposal that favors the SAC mechanism.[26,27] Our results, however, suggest SAC is associated with too high a barrier for hydrolysis. To examine the general base catalysis (GBC) mechanism involving E631, we also performed calculations following the same computational procedures described in this paper. The calculated rating limiting barrier for the E631-catalyzed GBC mechanism is 36.0 kcal/mol (see SI.7 for results of the GBC mechanism), suggesting that E631 is unlikely the postulated catalytic base.

The calculated GAC mechanism, based on potential energies, however, cannot be used directly to explain the solvent kinetic isotope effect (KIE) reported for HlyB-NBD, which gives $k_{cat,H2O}/k_{cat,D2O}$ a value of 0.79;[3] this inversed KIE appears to imply that the proton abstraction from water is not rate limiting. Inclusion of entropic contribution and nuclear quantum effects in calculations seem to be important to reconcile these kinetic data. Free energy simulations are currently underway to accurately determine the relative free energy barriers in GAC mechanism.

Protein and solvent assisted proton transfers have been discussed for other ATP hydrolyzing enzymes, such as $F_1$-ATPase,[28] PcrA Helicase,[29] DNA Polymerase IV,[30] and Myosin II,[31] where either an additional water (in the first three cases) or a Ser residue (Ser236) were proposed to mediate the proton transfer from the lytic water to one of the γ-phosphate oxygens via a relay mechanism. The reoccurrence of this mechanistic feature led us to speculate whether the assisted proton transfer represents a common theme in enzyme mediated ATP hydrolysis. We hypothesized that ATP-hydrolyzing enzymes might have evolved to incorporate a functionally equivalent residue in their active sites to facilitate proton transfer from the lytic water to ATP, which is intrinsically a slow step in solution and would become rate determining if not accelerated in enzymes. The catalytic motif that meets the chemical requirements to fulfill the proton relay are expected to contain residues that can provide an exchangeable proton at an effective $pK_a$ value (either intrinsically or modified by specific protein environment) between those of ATPH[3-] (~6.5) and water (~15.7). Further tests of this hypothesis may deepen our understandings of the common mechanism under which motor ATPases that utilize the RecA-like fold work.[32]

In summary, combined QM/MM calculations presented in this Communication revealed a novel catalytic role of the conserved H-loop residue H662 in the *E. coli* ABC transporter HlyB. Calculated reaction energy profiles indicate that H662 in HlyB-NBDs can catalyze hydrolysis of ATP by providing a "chemical linchpin" for proton relay such that the overall barrier height is greatly lowered. This proposal represents a new interpretation of the function of the conserved H-loop residue in ABC-transporters, as it differs significantly from the original view of this motif as an inert "linchpin" that merely holds the active residues together mechanically.

**Supporting Information**. Additional simulation details, key geometric features of stationary points, and movies of reaction paths identified from two dimensional potential energy surfaces of the GAC and the SAC mechanics are included in the supporting information (SI).

### Corresponding Author

*jpu@iupui.edu

### ACKNOWLEDGMENT

We thank GM Blackburn for comments on the manuscript and Nigel Richard and Wei Yang for helpful discussion. This work was supported by a start-up grant from Indiana Univ.-Purdue Univ. Indianapolis. The computing time is partly provided by Indiana University BigRed High Performance Computing, funded by National Science Foundations (NSF).

# Supporting Information

# H-loop Histidine Catalyzes ATP Hydrolysis in the *E. coli* ABC Transporter HlyB

*Yan Zhou, Pedro Ojeda-May, and Jingzhi Pu\**

Depart of Chemistry and Chemical Biology, Indiana University-Purdue University Indianapolis, Indianapolis, IN 46202

\* To whom correspondence should be addressed. Email: <u>jpu@iupui.edu</u>

## 1.  Schematic representation of hydrogen bonding interactions near the center of the active site in HlyB

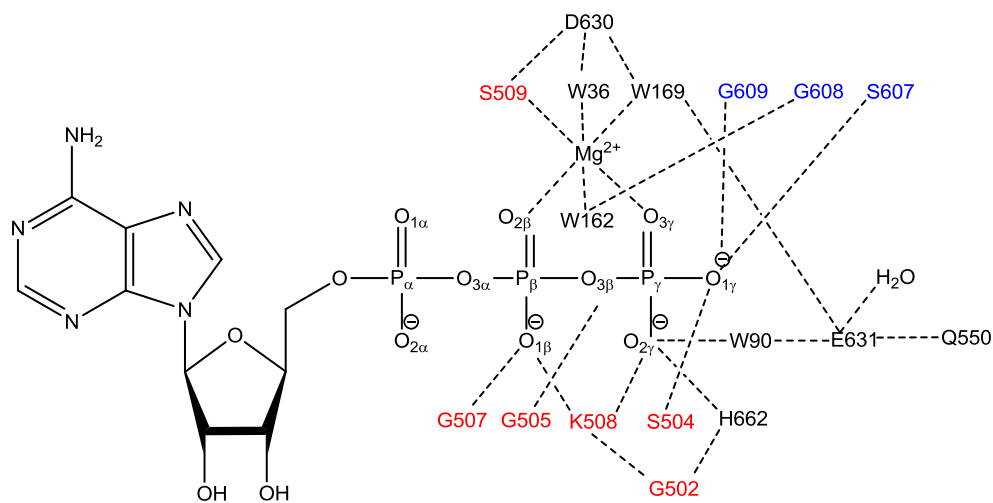

Scheme S1. Selected hydrogen bonding interactions in the active site of *E. coli* HlyB-NBDs. Residues in red are in the Walker A motif (or P-loop). Residues in blue are in the C-loop (or signature loop), which is from the opposite monomer. D630 is in the Walker B motif. E631 is near the end of Walker B. H662 is in the H-loop.

## 2. Details of building the simulation model by combing two HlyB structures.

The 1XEF structure contains ATP/$Mg^{2+}$ bound HlyB-NBDs with H662A mutations. The 2FGK structure contains E631Q mutations of HlyB-NBDs bound with ATP in the absence of $Mg^{2+}$. The simulation system was built based on the NBD dimer in the 1XEF structure, except that the 2FGK structure was used to model H662, E631, and the lytic water ($W_L$) (Figure 1b in the main text). The modeling of E631 is necessary because its original orientation in the 1XEF structure is not suited to activate the lytic water. The unsuitable side chain orientation of E631 may be caused by the mutation at the position of H662A in 1XEF. In contrast, the side chain orientation of Q631 in 2FGK, although presented as a mutant of E631, adopts a position that better aligns $W_L$ for an in line attack. Each of the two NBDs in the 2FGK structure contains a water molecule in the vicinity of $P_\gamma$ of ATP, H662, and Q631. Thus the positions of these crystal water molecules in 2FGK were also used to model $W_L$ in the 1XEF-based simulation system.

## 3. Justification of using the AM1/CHARMM method

The AM1 method is chosen here because of its stability, efficiency, and reasonable accuracy for the overall reaction energy based on the hydrolysis of methyl triphosphate in the gas phase as a model reaction (Pu *et al.*, unpublished). The underlying assumption of this choice is that systematic errors associated with the AM1 Hamiltonian would largely cancel out in distinguishing the operative mechanisms from those non-operative ones, especially when the same simulation protocol is applied to the same enzyme configuration. Note that the comparison of the GAC and the SAC mechanism in elucidating H662's functional role is the main focus of the present work.

## 4. Detailed description of the system preparation for the QM/MM calculations and additional SHAKE constraints

The protons in the guanidinium group of K508 were also constrained by SHAKE during MD simulations and energy scans to avoid irrelevant proton transfers. In addition, $H_N$ of H662 was constrained by SHAKE in the SAC mechanism.

## 5. Detailed discussions on geometries of stationary points for the GAC and the SAC mechanisms

### 5.1. Geometries and energetics of stationary points for the GAC mechanism

The PES of the GAC mechanism (Figure 2a) reveals a four-step process for ATP hydrolysis. Key geometric parameters and energetics of the stationary points determined for the GAC mechanism are listed in Table S1.

Table S1. Bond distances, dihedral angles, and energies of stationary points in the GAC mechanism (Figure 2a).[a]

|  | RC1 | RC2 | Distance (Å) | | | | | | $O_{1\gamma}$-$P_\gamma$-$O_{2\gamma}$-$O_{3\gamma}$ (degrees) | $\Delta E$ |
|  |  |  | $H_N$-$N_\varepsilon$ | $H_N$-$O_{2\gamma}$ | $H_W$-$O_W$ | $H_W$-$N_\varepsilon$ | $P_\gamma$-$O_{3\beta}$ | $P_\gamma$-$O_W$ |  |  |
|---|---|---|---|---|---|---|---|---|---|---|
| Min1 | -3.60 | -2.10 | 1.00 | 2.07 | 0.97 | 3.49 | 1.65 | 3.75 | 146 | -2.4 |
| TS1 | -2.90 | -1.19 | 1.00 | 2.04 | 0.96 | 2.82 | 1.65 | 2.85 | 150 | 9.0 |
| Min2 | -3.00 | -0.30 | 1.01 | 1.93 | 0.97 | 3.05 | 1.70 | 2.00 | 160 | 0.0 |
| TS2 | -1.81 | -0.30 | 1.11 | 1.47 | 0.99 | 2.44 | 1.71 | 2.01 | 162 | 15.0 |
| Min3 | -1.20 | -0.30 | 1.87 | 0.97 | 0.96 | 3.06 | 1.65 | 1.94 | 161 | 5.6 |
| TS3 | 2.30 | 0.29 | 3.68 | 0.95 | 1.11 | 1.54 | 2.02 | 1.72 | -169 | 22.1 |
| Min4 | 3.40 | 0.30 | 3.24 | 0.95 | 2.11 | 1.00 | 1.93 | 1.63 | -163 | -9.2 |
| TS4 | 3.40 | 1.50 | 3.34 | 0.96 | 2.03 | 1.00 | 3.09 | 1.59 | -134 | 4.7 |
| Min5 | 3.40 | 1.70 | 3.34 | 0.96 | 2.03 | 1.01 | 3.29 | 1.59 | -131 | 3.1 |

[a]Distances are in Å, dihedral angles in degrees, and energies in kcal/mol.

In the first step of the calculated GAC mechanism, the catalytic water approaches ATP to form a reactant complex Min1. The second step of the GAC mechanism is a proton transfer from H662 to the electronegative $O_{2\gamma}$ of ATP with a barrier height of 15.0 kcal/mol (TS2). The energy of the resulting intermediate (Min3) is 5.6 kcal/mol relative to the reactant state (Min2). Hydrogen bonding interactions involving the imidazolate ion in Min3 include a bulk water molecule, D637, G502, and A635, which may contribute to the stability of this intermediate. In TS2, the $H_N$-$N_\varepsilon$ and the $H_N$-$O_{2\gamma}$ bonds are 1.47 Å and 1.11 Å, respectively, indicating that the H662 proton $H_N$ is being transferred to ATP γ-phosphate. The proton transfer from H662 to ATP is decoupled from phosphoryl transfer, as the $P_\gamma$-$O_{3\beta}$ bond does not change significantly (the $P_\gamma$-$O_{3\beta}$ distance is 1.70, 1.71 and 1.65 Å in Min2, TS2, and Min3 respectively), and no $P_\gamma$-$O_w$ bond formation is detected (the $P_\gamma$-$O_w$ distance is 2.00, 2.01, and 1.94 Å in Min2, TS2, and Min3 respectively). While the $H_W$-$N_\varepsilon$ bond distance varies from 3.05 Å in Min2, through 2.44 Å in TS2, to 3.06 Å in Min3, the $H_W$-$O_W$ bond is almost unchanged during the transition, suggesting a stepwise proton relay, where H662's donating and accepting proton are well separated. The third step consists of two concerted processes in which the proton transfer of $H_W$ from the lytic water $W_L$ to $N_\varepsilon$ of H662 is coupled to the $P_\gamma$-$O_W$ bond formation, followed by breaking of the $P_\gamma$-$O_{3\beta}$ bond (Scheme 1). As the rate limiting step in the GAC mechanism, the third step goes through a transition state TS3 that is associative in nature and gives rise to an overall barrier height of 22.1 kcal/mol. In TS3, the $H_W$-$O_W$ bond is elongated to 1.11 Å and the $H_W$-$N_\varepsilon$ bond is shortened to 1.54 Å. The $P_\gamma$-$O_{3\beta}$ bond is stretched to 2.02 Å and the $P_\gamma$-$O_W$ bond is almost formed at a bond length of 1.72 Å. In the resulting intermediate Min4, the $H_W$-$N_\varepsilon$ bond (1.0 Å) and the $P_\gamma$-$O_W$ bond (1.63 Å) are completely formed, but the $P_\gamma$-$O_{3\beta}$ bond has not broken yet at this point (1.93 Å). In both TS3 and Min4, $P_\gamma$ has a trigonal bipyramidal bonding pattern. In Min4, the dihedral angle $O_{1\gamma}$-$P_\gamma$-$O_{2\gamma}$-$O_{3\gamma}$ is -163° and the angle $O_{3\beta}$-$P_\gamma$-$O_W$ is 166° (not shown in Table S1). Min4 is the most stable state in the whole reaction path (-9.2 kcal/mol). In the last step, cleavage of the $P_\gamma$-$O_{3\beta}$ bond leads to the final product complex (Min5), with a $P_\gamma$-$O_{3\beta}$ bond of 3.29 Å.

E631 always binds with $W_L$ in the whole process and may serve to stabilize the active site in this mechanism. The length of the hydrogen bond between the sidechain of E631 and $W_L$ ranges from 1.62 to 2.19 Å. The hydrogen bond between E631 and $W_L$ is about 0.5 Å shorter in Min5 than in Min4, which shows that E631 may assist in the $P_\gamma$-$O_{3\beta}$ bond breaking. The close interaction between E631 and Pi suggests that E631 may act as a mechanical hinge to the rotation of the helical domain and NBD-TMD communication (Zaitseva *et al. EMBO J* **2006**, *25*, 3432). K508 hydrogen bonds with $O_{1\beta}$ and $O_{2\gamma}$ of ATP simultaneously. The lengths of these two hydrogen bonds are almost equivalent most of the time. In TS3, K508 moves closer to $O_{2\gamma}$ to better stabilize the transition state. During the transition from Min4 to Min5, when Pi moves away from ADP, K508 retains the hydrogen bond with $O_{1\beta}$ of ADP; in contrast, the hydrogen bond between K508 with $O_{2\gamma}$ is weakened, but still within the a reasonable hydrogen bond distance range. In Min5, the H-loop residue H662 forms hydrogen bonds with both $O_{2\gamma}$ and $O_W$ of Pi. Two Ser residues, i.e., the P-loop S504 and the signature loop S607 (from the opposite monomer) form hydrogen bonds with $O_{1\gamma}$ throughout the whole simulation of the reaction. In Min5, the Pi moiety continues to bind to $Mg^{2+}$ ion after hydrolysis, making spontaneous Pi release from the active site less likely. The crystal water W90 forms hydrogen bonds with $O_{2\gamma}$ from Min1 to Min3, and then with $N_\varepsilon$ in TS3. The other hydrogen of W90 binds to one of E631 carboxylate oxygens throughout the simulation.

## 5.2. Geometries and energetics of stationary points for the SAC mechanism

The PES obtained for the SAC mechanism (Figure 2b) also displays a stepwise pattern for ATP hydrolysis. Key geometric parameters and energetics of the stationary points determined for the SAC mechanism are listed in Table S2.

Table S2. Bond distances, dihedral angles, and energies of stationary points in the SAC mechanism (Figure 2b).[a]

|  | RC1 | RC2 | Distance (Å) | | | | $O_{1\gamma}$-$P_\gamma$-$O_{2\gamma}$-$O_{3\gamma}$ (degrees) | $\Delta E$ |
|  |  |  | $H_W$-$O_W$ | $H_W$-$O_{2\gamma}$ | $P_\gamma$-$O_{3\beta}$ | $P_\gamma$-$O_W$ |  |  |
| Min1 | -2.30 | -2.10 | 0.97 | 3.28 | 1.66 | 3.76 | 148 | -6.4 |
| TS1 | -2.10 | -1.20 | 0.97 | 3.08 | 1.65 | 2.85 | 150 | 7.7 |
| Min2 | -1.60 | -0.30 | 0.96 | 2.55 | 1.68 | 1.98 | 159 | 0.0 |
| TS2 | -0.10 | -0.10 | 1.20 | 1.31 | 1.72 | 1.83 | 170 | 32.1 |
| Min3 | 1.20 | 0.10 | 2.15 | 0.95 | 1.77 | 1.68 | -173 | -0.8 |
| TS3 | 2.19 | 0.29 | 3.14 | 0.95 | 1.97 | 1.68 | -162 | 9.8 |
| Min4 | 2.29 | 0.30 | 3.24 | 0.95 | 1.94 | 1.65 | -160 | -6.7 |
| TS4 | 2.49 | 1.40 | 3.45 | 0.96 | 3.01 | 1.62 | -135 | 9.1 |
| Min5 | 2.49 | 1.80 | 3.46 | 0.97 | 3.40 | 1.61 | -129 | 7.8 |

[a]Distances are in Å, dihedral angles in degrees, and energies in kcal/mol.

The SAC reaction route is initiated by formation of a reactant complex (Min2) between $W_L$ and ATP, in which the distance between $P_\gamma$ and $O_W$ is reduced to 1.98 Å. The second step in SAC is rate limiting and involves two concerted processes that couple proton transfer between the lytic water and $O_{2\gamma}$ of ATP to the $P_\gamma$-$O_W$ bond formation, by going through a trigonal bipyramidal transition state (TS2), in which both dihedral angle $O_{1\gamma}$-$P_\gamma$-$O_{2\gamma}$-$O_{3\gamma}$ and the angle $O_{3\beta}$-$P_\gamma$-$O_W$ are about 170°. The $H_W$-$O_W$ and $H_W$-$O_{2\gamma}$ bonds in TS2 are 1.20 and 1.31 Å, respectively; the $P_\gamma$-$O_W$ bond is shortened to 1.83 Å from compared to 1.98 Å in Min2, whereas the $P_\gamma$-$O_{3\beta}$ (1.72 Å) is slightly longer than that in Min2 (1.68 Å) and Min1 (1.66 Å). The completion of the proton transfer step in SAC leads to a phosphrane intermediate (Min3) with an energy of -0.8 kcal/mol. In Min3, the newly formed $H_w$-$O_{2\gamma}$ and $P_\gamma$-$O_W$ bonds are 0.95 and 1.68 Å, respectively, indicating that both proton transfer and $P_\gamma$-$O_W$ bond formation are completed. The $P_\gamma$-$O_{3\beta}$ bond in Min3 is stretched to 1.77 Å, only about 0.1 Å longer than that in Min1 and Min2, implying that the above steps are separated from the $P_\gamma$-$O_{3\beta}$ bond scission process next. The third step from Min3 to Min4 corresponds to a structural relaxation of hydrogen bonding interactions involving H662 and E631. This relaxation step is mainly due to mainly due to the rotation of the $O_{2\gamma}$-$H_W$ group about the $P_\gamma$-$O_{2\gamma}$ bond. As a result, the $O_{2\gamma}$-$H_W$ bond that originally points to H662 in Min3 is oriented toward the β phosphate in Min4, accompanied by a change of the dihedral angle $H_W$-$O_{2\gamma}$-$P_\gamma$-$O_{3\gamma}$ from -50° to 103°. In Min4, $H_W$ is ready to form a hydrogen bond with the increasingly negatively charged $O_{3\beta}$ atom due to the stretched $P_\gamma$-$O_{3\beta}$ bond distance. In addition, there is a hydrogen bond formed between the non-transferred hydrogen on $W_L$ and E631. The resulting intermediate Min4 is the most stable state on the reaction path. The final step in the SAC mechanism is the cleavage of the $P_\gamma$-$O_{3\beta}$ bond, resulting in a $P_\gamma$-$O_{3\beta}$ bond of 3.40 Å in Min5.

### 6. Movies.

Movie S1 shows the reaction path identified from Figure 2a for the GAC mechanism, where H662 assists a proton transfer from the lytic water to one of the $\gamma$-phosphate oxygens of ATP through a relay. Movie S2 shows the reaction path identified from Figure 2b for the SAC mechanism, where the proton on the lytic water is directly transferred to one of the $\gamma$-phosphate oxygens of ATP.

### 7. Simulations and results of the mechanism of using E361 as a general base and the detailed discussion

To examine the validity of the "catalytic carboxylate" proposal, also referred to as the general base catalysis (GBC) mechanism, we performed additional simulations for or HlyB mediated ATP hydrolysis using E631 as a catalytic base. Two-dimensional potential energy surface (PES) was obtained based on two reaction coordinates RC1 and RC2, where RC1 represents a combined proton transfer reaction coordinate describing proton transfer from lytic water to E631, i.e. $RC1^{GBC} = O_W\text{-}H_W - H_W\text{-}OE2_{E631}$, and RC2 is the reaction coordinate that describes phosphoryl transfer, i.e. $RC2^{GBC} = O_{3\beta}\text{-}P_\gamma - P_\gamma\text{-}O_W$. The 2d-PES of the GBC mechanism is shown in Fig. S1. Key geometries and energetics of stationary points along the reaction path identified for the GBC mechanism (using E631 as a general base) are listed in Table S3. Comparison of energy profiles from GBC, GAC, and SAC is shown in Fig. S2.

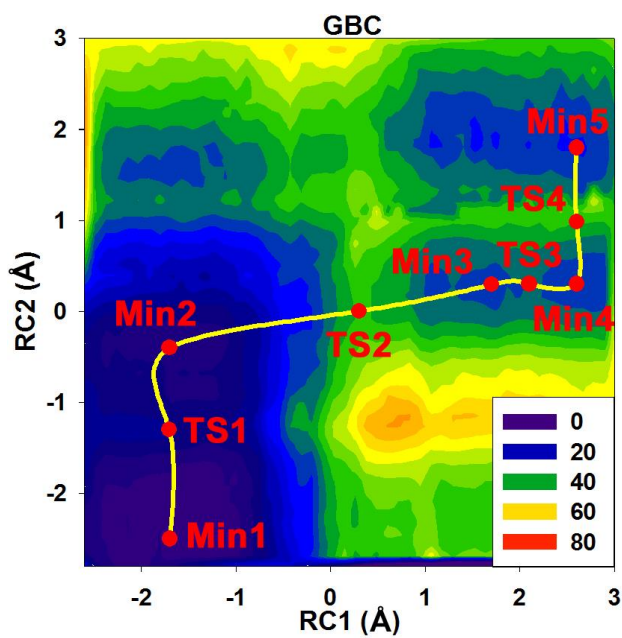

Figure S1. Two-dimensional potential energy surfaces of the general base catalysis (GBC) mechanism where the lytic proton is initially transferred to the general base E631 in the HlyB-NBD mediated ATP-hydrolysis. For the GBC mechanism, the abscissa (RC1) represents a proton-transfer reaction coordinate which is negative as the proton is bonded to the lytic water, and positive as the proton is transferred to the E631 carboxylate oxygen. The ordinate (RC2) is the antisymmetric combination of the $O_{3\beta}$-$P_\gamma$ and $P_\gamma$-$O_W$ bond distances, in describing the phosphoryl transfer reaction during ATP hydrolysis. Distances are in Å, and energies are in kcal/mol.

Table S3. Energies and geometries of stationary points for the GBC mechanism.[a]

| | RC1 | RC2 | Bond distances (Å) | | | | $O_{1\gamma}$-$P_\gamma$-$O_{2\gamma}$-$O_{3\gamma}$ (degrees) | $\Delta E$ |
| | | | $H_W$-$O_W$ | $H_W$-OE2 | $P_\gamma$-$O_{3\beta}$ | $P_\gamma$-$O_W$ | | |
|---|---|---|---|---|---|---|---|---|
| Min1 | -1.70 | -2.50 | 0.96 | 2.66 | 1.66 | 4.16 | 150 | -5.4 |
| TS1 | -1.71 | -1.30 | 0.97 | 2.67 | 1.70 | 3.00 | 155 | 6.9 |
| Min2 | -1.70 | -0.40 | 0.97 | 2.67 | 1.69 | 2.09 | 161 | 0.0 |
| TS2 | 0.30 | 0.00 | 1.40 | 1.10 | 1.85 | 1.84 | -179 | 36.0 |
| Min3 | 1.70 | 0.30 | 2.68 | 0.98 | 2.03 | 1.73 | -167 | 25.5 |
| TS3 | 2.10 | 0.30 | 3.08 | 0.98 | 2.03 | 1.73 | -167 | 26.8 |
| Min4 | 2.60 | 0.30 | 3.58 | 0.98 | 2.04 | 1.74 | -166 | 24.8 |
| TS4 | 2.60 | 0.99 | 3.58 | 0.98 | 2.63 | 1.64 | -149 | 34.0 |
| Min5 | 2.60 | 1.80 | 3.58 | 0.98 | 3.46 | 1.66 | -142 | 20.6 |

[a]Distances are in Å, dihedral angls in degrees, and energies in kcal/mol.

Similar to GAC and SAC, the first step of the GBC mechanism is associated with the formation of a reactant complex. The second step of in the GBC mechanism is a concerted reaction where the transfer of $H_W$ from $O_w$ of $W_L$ to OE2 of E631 and the $P_\gamma$-$O_W$ bond formation occur simultaneously. In the pentacovalent transition state TS2 for this step, $H_W$-$O_W$ and $H_W$-OE2 are 1.40 and 1.10 Å, respectively, and the bond lengths of $P_\gamma$-$O_{3\beta}$ and $P_\gamma$-$O_W$ are 1.85 and 1.84 Å, respectively. In Min3, which corresponds to a phosphorane-like intermediate, although Pi has not been fully cleaved from ATP, the bond between $P_\gamma$ and ADP is weakened as reflected from an elongated $P_\gamma$-$O_{3\beta}$ bond of 2.03 Å, meanwhile the $P_\gamma$-$O_W$ bond has been formed (1.73 Å). The transition state TS3 mainly corresponds to relaxation involving the protonated E631 after the $H_W$ transfer step; the distance between $O_w$ and E631 increases during the relaxation, indicating weakened hydrogen bonds after E631 is pronated. The transition state TS4 in the last step is associated with breaking the $P_\gamma$-$O_{3\beta}$ bond, leading to formation of the final product complex (Min5).

E631 hydrogen bonds with both $W_L$ and the backbone hydrogen of H662 before Min2. From Min2 to TS2, the hydrogen bond between E631 and H662 increases gradually from 2.16 Å to 4.48 Å so that E631 moves closer to $W_L$ to accept $H_W$. Then this hydrogen bond is weakened and adopts a distance of 2.45 Å in Min3 along with the separation of E631 from $W_L$. In Min5, this hydrogen bond is shortened to 1.93 Å, lowering the energy of the system. The crystal water W90 stabilizes both $O_{2\gamma}$ and E631 in the initial reactant state structures. After the proton transfer from $W_L$ to E631, the hydrogen bond between W90 and E631 is also weakened. Along the whole reaction process, H662 always maintains hydrogen bond interactions with $O_{2\gamma}$. The crystal water W109 is in the vicinity of $W_L$ but it has no obvious hydrogen bond with $W_L$ prior to TS2. After TS2 is formed, W109 approaches and binds to $O_W$ with a hydrogen bond of 2.31 Å in Min3. The interaction between W109 and $O_W$ is stronger in Min5, showing that in the GAC mechanism, this water helps the $P_\gamma$-$O_{3\beta}$ bond breaking instead of E631 as in the GAC and SAC mechanisms.

Figure S2 shows the energy profile of the GBC mechanism (blue) compared with those from GAC (black) and SAC (red). One can see that the GBC mechanism has a barrier height of 36.0 kcal/mol for the rate determining step, corresponding to the proton transfer from the lytic water to carboxylate of E631. This result suggests that the GBC mechanism involving E631 is not an operative one compared to the GAC mechanism in which proton relay through H-loop H662 can be utilized to lower the overall barrier height of ATP hydrolysis by HlyB-NBDs.

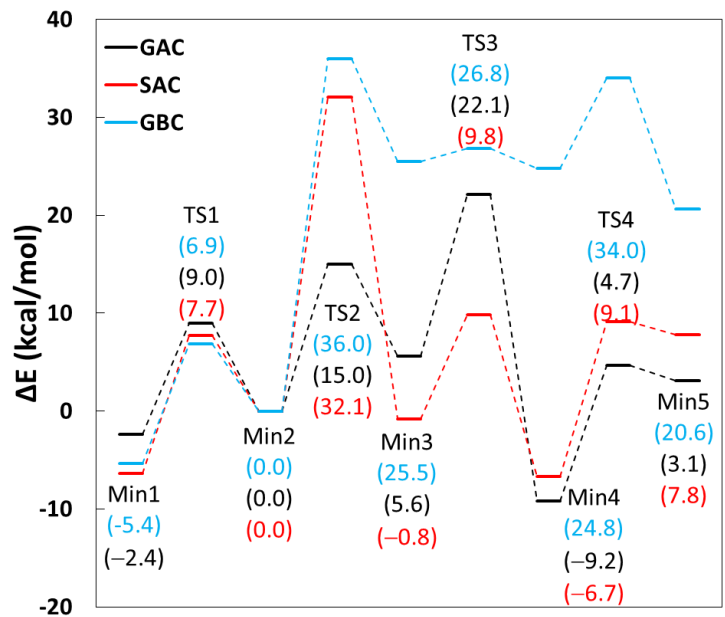

Figure S2. Comparison of energy profiles from the GBC, GAC, and SAC mechanisms.